# Noncollinear Spintronics and Electric-Field Control: A Review

Peixin Qin, Han Yan, Xiaoning Wang, Zexin Feng, Huixin Guo, Xiaorong Zhou, Haojiang Wu, Xin Zhang, Zhaoguogang Leng, Hongyu Chen, Zhiqi Liu*

School of Materials Science and Engineering, Beihang University, Beijing 100191, China

Email: zhiqi@buaa.edu.cn

**Abstract:** Our world is composed of various materials with different structures, where spin structures have been playing a pivotal role in spintronic devices of the contemporary information technology. Apart from conventional collinear spin materials such as collinear ferromagnets and collinear antiferromagnetically coupled materials, noncollinear spintronic materials have emerged as hot spots of research attention owing to exotic physical phenomena. In this Review, we firstly introduce two types noncollinear spin structures, *i.e.*, the chiral spin structure that yields real-space Berry phases and the coplanar noncollinear spin structure that could generate momentum-space Berry phases, and then move to relevant novel physical phenomena including topological Hall effect, anomalous Hall effect, multiferroic, Weyl fermions, spin-polarized current, and spin Hall effect without spin-orbit coupling in these noncollinear spin systems. Afterwards, we summarize and elaborate the electric-field control of the noncollinear spin structure and related physical effects, which could enable ultralow power spintronic devices in future. In the final outlook part, we emphasize the importance and possible routes for experimentally detecting the intriguing theoretically predicted spin-polarized current, verifying the spin Hall effect in the absence of spin-orbit coupling and exploring the anisotropic magnetoresistance and domain-wall-related magnetoresistance effects for noncollinear antiferromagnetic materials.

## 1. Introduction

Electrons have two intrinsic attributes, *i.e.*, charge and spin. In the last century, traditional semiconductor electronic devices were achieved by manipulating the motion of electrons which utilized the electron charge. With the rapid development of microelectronics, the increasing demand for the technique of high-density and low-power storage made people start to focus on another property of electrons - spin. In 1988, Albert Fert and Peter Andreas Grünberg simultaneously discovered the giant magnetoresistance effect [1, 2], opening the door of spintronics. Controlling the spin of electrons effectively and precisely is the core of spintronics which can realize fast-response-speed, low-energy-loss, non-volatile spintronic devices. Subsequently, the discovery of tunnel magnetoresistance, spin transfer torque and spin orbit torque further motivated the rapid development of spintronics [3-13].

As the spins of electrons could orient in any direction, the structure of spins could be miscellaneous. Here, we generally classify the spin arrangement into two kinds of patterns, collinear spin and noncollinear spin structures. Materials with a noncollinear spin structure can be ferromagnets, ferrimagnets and antiferromagnets. Moreover, we can further classify the noncollinear spin structure into noncoplanar and coplanar noncollinear structures. In recent years, the study on the noncollinear spin structure has drawn great attention because of various novel physical phenomena such as skyrmions, the topological Hall effect, the multiferroic effect, the anomalous Hall effect in antiferromagnets, spin-polarized current in antiferromagnets, Weyl fermions and the possible spin Hall effect in the absence of spin-orbit coupling as summarized in Fig. 1.

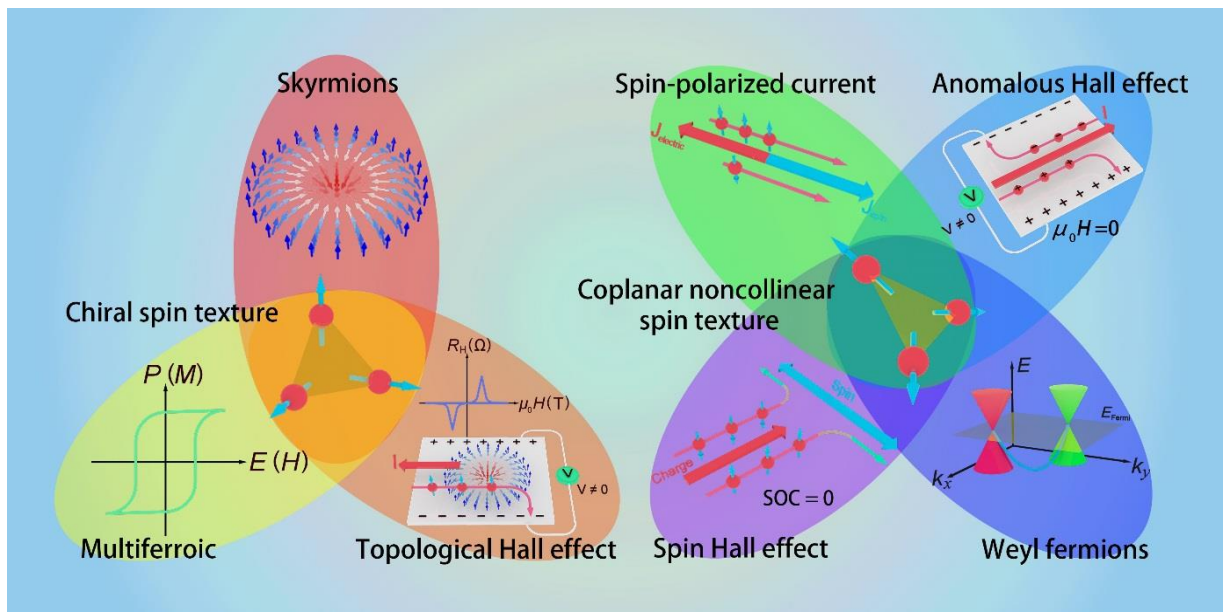

**Fig.1** Schematic of structures and physical phenomena in chiral and coplanar noncollinear spin textures.

The spin degree of freedom can be modulated by electric current and it plays an important role in the information storage technique such as hard disks. Nevertheless, the current gives rise to a large amount of Joule heat and thus cause energy consumption. Nowadays, tuning the electron spin by electric field instead of current is a high-efficiency method which could avoid useless energy loss. In our previous work, we have imposed the electric-field manipulation in collinear magnetic materials, thus tailoring the magnetic phase transitions in ferromagnetic FeRh [14] and demonstrating a collinear antiferromagnetic memory device based on MnPt [15]. In this Review, we will first introduce the exotic structure and attractive physical effects in noncollinear spin materials. Subsequently, electric-field control of the noncollinear spintronics will be presented.

## 2. Noncollinear spin structure and physical phenomena

### 2.1 Skyrmions and topological Hall effect

In 1962, Skyrme [16] proposed a nonlinear field theory for the interactions of pions and predicted a topologically protected particle-like object with field-stable structure which has been named "skyrmion". Later in 1975 [17], Belavin *et al.* theoretically predicted a similar particle-like metastable spin structure existing in two-dimensional ferromagnets. Skyrmions possess a chiral spin structure and a vortex configuration and they are thought to be small swirling topological defects in the magnetization texture. The neighboring moments orient perpendicularly to each other under the influence of Dzyaloshinskii–Moriya interaction [18, 19] that originates from inversion symmetry breaking and spin-orbit coupling. Besides, the Dzyaloshinskii–Moriya interaction can reduce the energy in skyrmions and make them remain stable in the magnetic field. There are two types of skyrmions: Bloch-type skyrmion usually occurring in bulk materials and Néel-type skyrmions typically arising in thin films.

In 2006, Rößler *et al.* [20] theoretically identified that skyrmions could spontaneously and stably exist in non-centrosymmetric materials in the absence of inversion symmetry. Additionally, the candidate bulk materials for realizing this novel state are particularly referred to the cubic *B*20 crystal structure, for example, FeGe and MnSi. Three years later in 2009, skyrmions were observed in bulk MnSi [21-23]. Later, skyrmions were discovered in ultrathin magnetic films epitaxially grown on heavy metals in 2011[24]. Yu *et al.* [25] for the first time reported the formation of a skyrmion crystal in FeGe near room temperature (Fig. 2). Additional studies by Ohuchi *et al.* [26] found skyrmions in centrosymmetric EuO thin films, implying that skyrmions are not only restricted to chiral materials or interfaces but varieties of formation mechanisms such as long-ranged magnetic dipolar interactions [27-29], frustrated exchange interactions [30] and four-spin exchange interactions [24] could work.

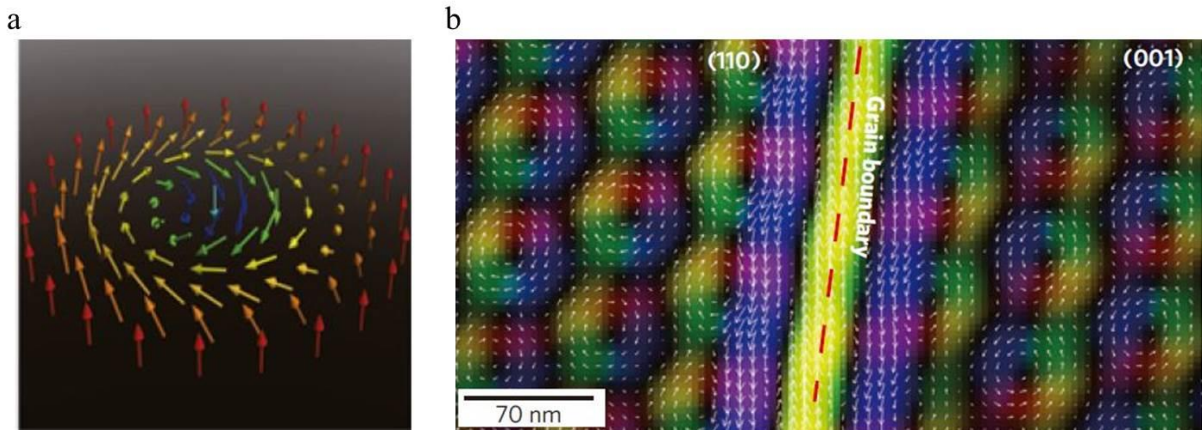

**Fig.2 a** Sketch of a skyrmion with chiral spin structure and **b** distribution of magnetization in (110) and (001) plane in FeGe.[25]

In recent years, the motion of skyrmions driven by spin orbit torque or spin transfer torque makes them promising candidates for spintronic devices. Yet, the size of skyrmions generated from the Dzyaloshinskii–Moriya interaction is ~100 nm, which mismatches the crucial nanoscale for the practical high-density device applications. In addition, skyrmions typically exists in a certain temperature and magnetic field range. Thus, how to attain a stable nanoscale skyrmion under room temperature and a small applied field is a crucial task for device applications. Recently, Legrand *et al.* [31] utilized magnetic force microscopy to image the skyrmions in the synthetic antiferromagnets at room temperature (Fig. 3). In contrast to the skyrmions in ferromagnetic materials, the skyrmions in antiferromagnetic materials can obtain a smaller size and more efficient motion. Moreover, the stability at room temperature makes it potential for low-consumption spintronic devices

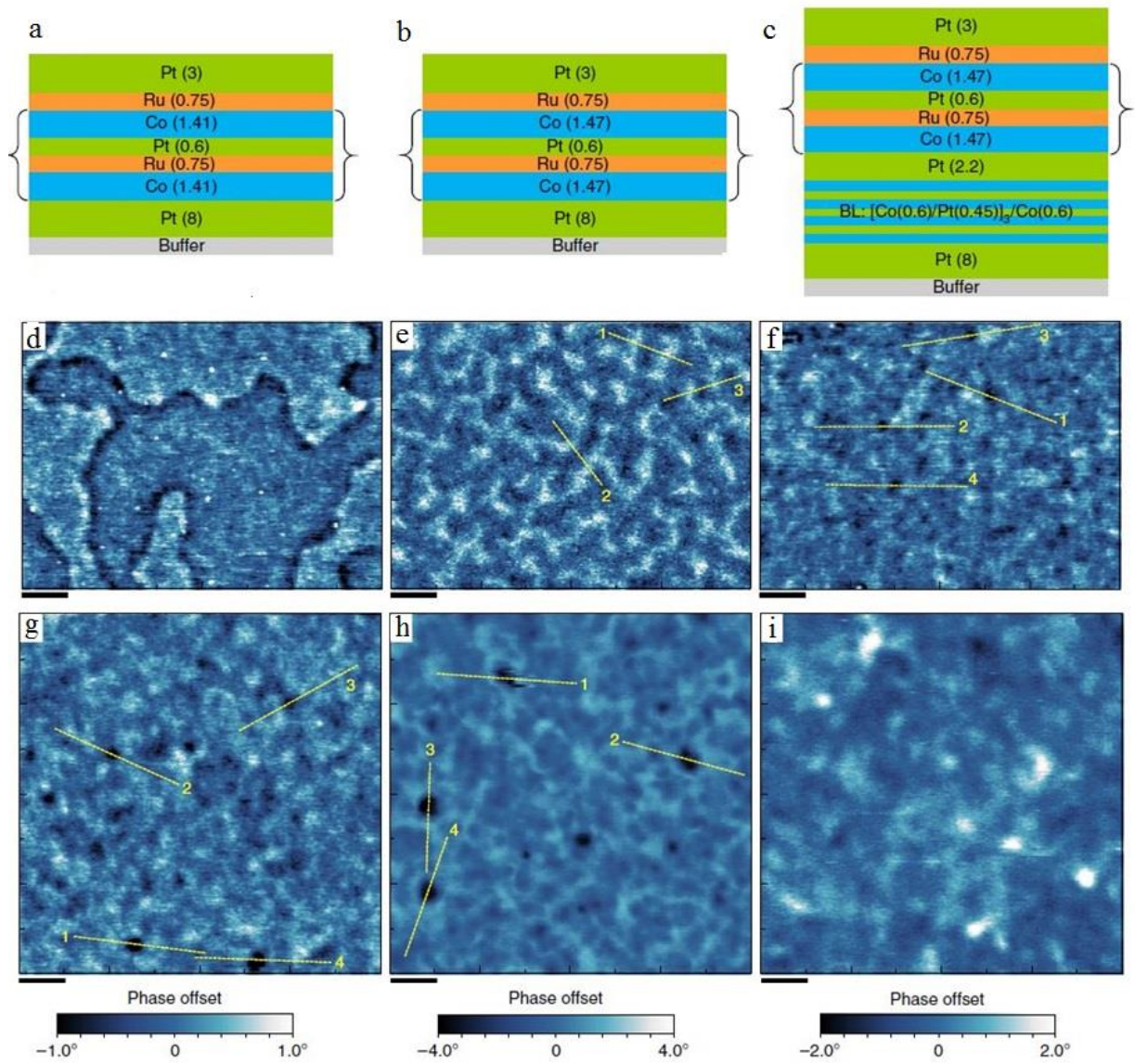

**Fig.3** Schematics of multilayer structures for synthetic antiferromagnets **a** possessing effective perpendicular magnetic anisotropy (PMA), **b** with a vanishing effective PMA, and **c** combining a vanishing effective PMA and a bias layer (BL). **d-f** Magnetic force microscopy (MFM) images of **a, b** and **c** at 0 mT, respectively. MFM observations of the BL-SAF system under external applied perpendicular field for $\mu_0 H_{ext} =$ **g** 20 mT, **h** 60 mT and **i** 100 mT. [31]

Unusual bulge and dent in the Hall resistance curves as a function of magnetic field are considered to be the features of the topological Hall effect, which stems from the scalar spin chirality in real space and is extensively utilized to detect the presence of chiral spin structure such as skyrmions. In 2009, Neubauer *et al.* [21] reported that the topological Hall effect in the *A* phase of the transition metal compound MnSi, which testified the skyrmions structure with topological properties (Fig. 4a and b). Their work established a direct connection between the topological Hall effect and the chiral spin structure.

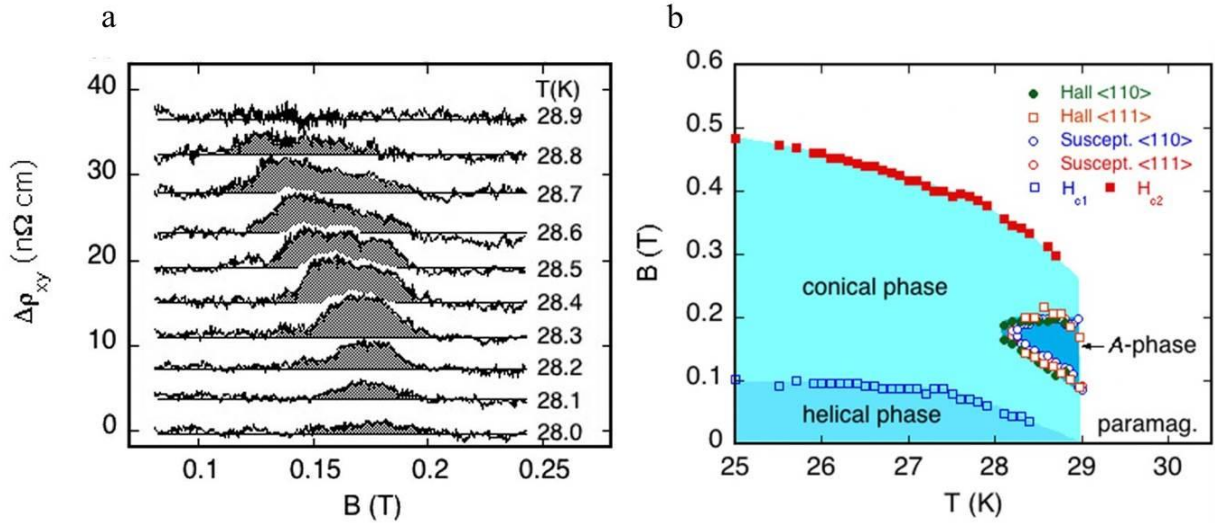

**Fig.4 a** Anomalous Hall resistivity in *A* phase and **b** magnetic phase diagram of MnSi. [21]

In addition to the discovery of the topological Hall effect and skyrmions in intermetallic compounds, Matsuno *et al.* [32] revealed the topological Hall effect in epitaxial bilayers composed of ferromagnetic oxide $SrRuO_3$ and paramagnetic $SrIrO_3$ over a large range of magnetic field and temperature. As shown in Fig. 5a, the topological Hall effect can be observed below 80 K and becomes more obvious while decreasing the temperature. In the high magnetic field, the Hall resistivity only comes from anomalous Hall effect owing to the lack of scalar spin chirality. As a result, the anomalous Hall effect and topological Hall effect can be separated by undertaking the Kerr rotation angle. The topological Hall conductivity is ~0.2 μΩ·cm with the same scale of the anomalous Hall conductivity (Fig. 5b). Both the switching magnetic field and peak field decline as the temperature rises (Fig. 5c). More importantly, the size of skyrmions induced by the interfacial Dzyaloshinskii–Moriya interaction can reach ten nanometers, which can satisfy the demand of nanoscale spintronic devices. Furthermore, Wang *et al.* [33] recently discovered the topological Hall effect in the single-layer ferromagnetic ultra-thin films, SrRuO$_3$ and V-doped Sb$_2$Te$_3$ (Fig. 6), and they suggested that the spin chiral might be a common feature of two-dimensional ultra-thin ferromagnetic materials with perpendicular magnetization.

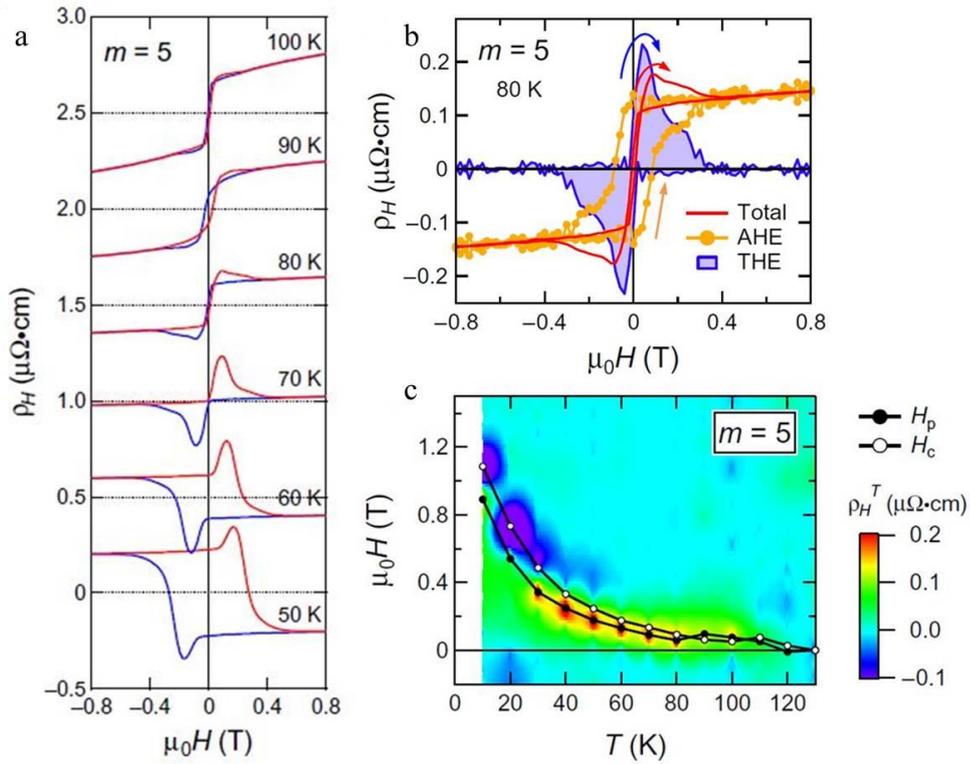

**Fig.5 a** Anomalous Hall resistivity of (SrRuO$_3$)$_m$-(SrIrO$_3$)$_2$ bilayers from 50 to 100 K at m = 5 . **b** Respective contribution from topological Hall effect and anomalous Hall effect and **c** Contour plot of anomalous Hall resistivity with coercivity and peak position. [32]

**2.2 Anomalous Hall effect and Weyl fermions in noncollinear antiferromagnets**

In 1879, Edwin Hall initially presented that when a charge current flow through a conductor, a voltage would generate in the vertical direction because the electrons mobile transversely under the applied field [34]. Two years later, he further found a much larger effect in ferromagnetic materials which was coined as the anomalous Hall effect [35]. As the Lorentz force is not the dominant factor any more, the anomalous Hall effect had been thought to be proportional to magnetization. From this perspective, the anomalous Hall effect should be absent in antiferromagnetic materials with zero net magnetization.

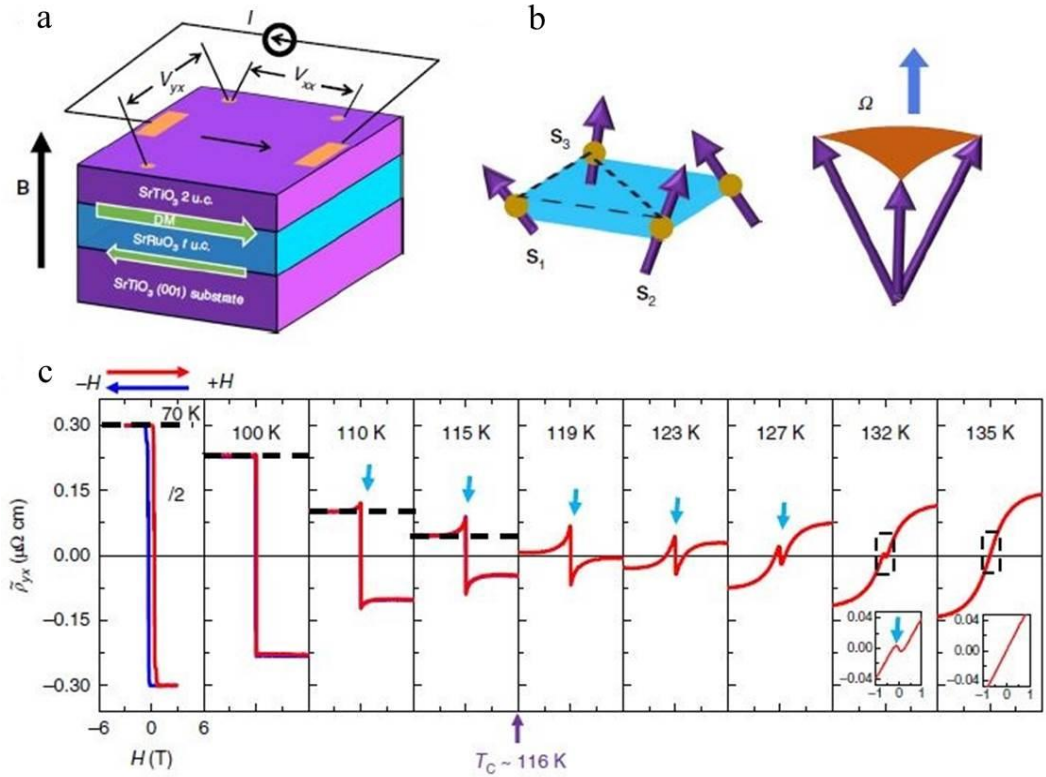

**Fig.6 a** Sketch of bilayers of 2.u.c. $SrTiO_3$ and t.u.c. $SrRuO_3$ grown on $SrTiO_3$ substrates. **b** Chiral spin structure in $SrRuO_3$ films. **c** Hall resistivity of $SrRuO_3$ films from 70 to 135 K. [33]

However, studies on the anomalous Hall effect in the recent two decades reveal that both the intrinsic mechanism Berry curvature and extrinsic mechanisms including skew scattering and side jump can yield anomalous Hall conductivity. As a result, in 2014, Chen *et al.* [36] theoretically predicted that a surprisingly large anomalous Hall effect that is the same order of magnitude with ferromagnetic transition metals exists in the noncollinear antiferromagnetic material $Mn_3Ir$. Analogy to $Mn_3Pt$ and $Mn_3Rh$, $Mn_3Ir$ crystallizes in a face-centered cubic lattice with Mn sublattices consisting of two-dimensional kagome lattices. As shown in Fig.

7a and b, the Berry curvature reaches maximum at the knots in the momentum space, which is the intrinsic origin for the anomalous Hall effect. Moreover, Fig. 7c depicts that the spin-orbit coupling is essential for the anomalous Hall effect. This work indicated that the anomalous Hall effect might originate from the combined influence of spin-orbit coupling and time reversal symmetry breaking through calculations. And they presumed the anomalous phenomenon could also occur in same-structure $Mn_3Pt$.

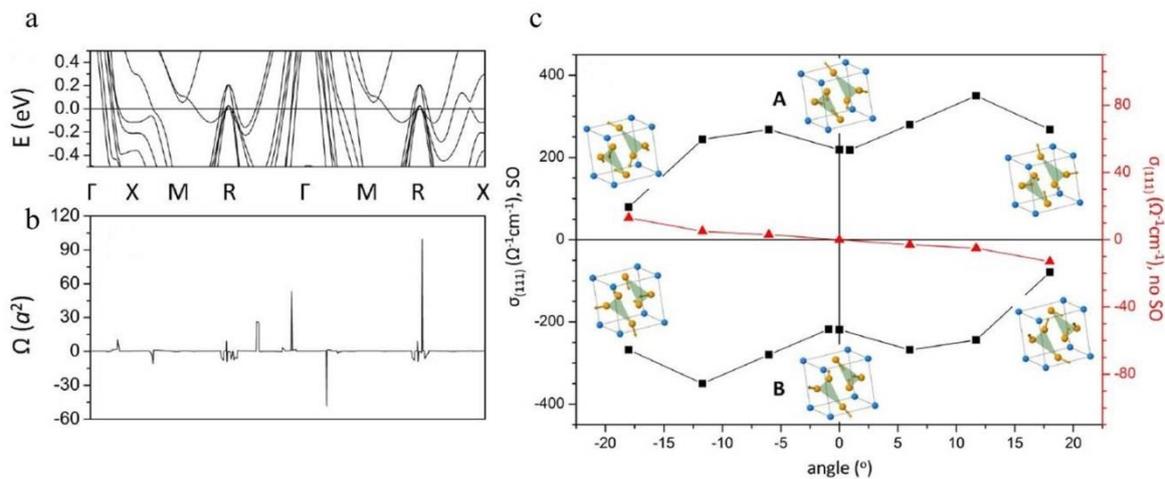

**Fig.7 a** Band structure of $Mn_3Ir$ near the Fermi level. **b** Berry curvature along the same k-point path as **a**. **c** Anomalous Hall conductivities with and without spin-orbit coupling under dependence with the tilt angle of Mn moment. [36]

Soon after the work above, Nakatsuji *et al*. [37] experimentally found an obvious anomalous Hall effect in the noncollinear antiferromagnetic $Mn_3Sn$ with triangular spin ordering. Figure 8a shows that the magnetization is ~6 m$\mu_B$/Mn at room temperature and the non-zero net magnetization is related to the spin canting in the noncollinear spin structure. The magnetization decreases with temperature and vanishes at ~450 K (Fig. 8a). Additionally, the anomalous Hall effect can be observed from 100 to 400 K and the Hall resistivity at room temperature is ~4 μΩ·cm (Fig. 8b). It is interesting to note that the anomalous Hall resistivity is not proportional to the magnetization, suggesting that the anomalous Hall effect is generated from Berry curvature rather than the weak net magnetization. The authors consider that the chiral spin structure breaks the in-plane hexagonal symmetry and thereby results in the non-zero Berry curvature. These results well verify the theoretical predictions on the large anomalous Hall effect in noncollinear antiferromagnets [36, 38].

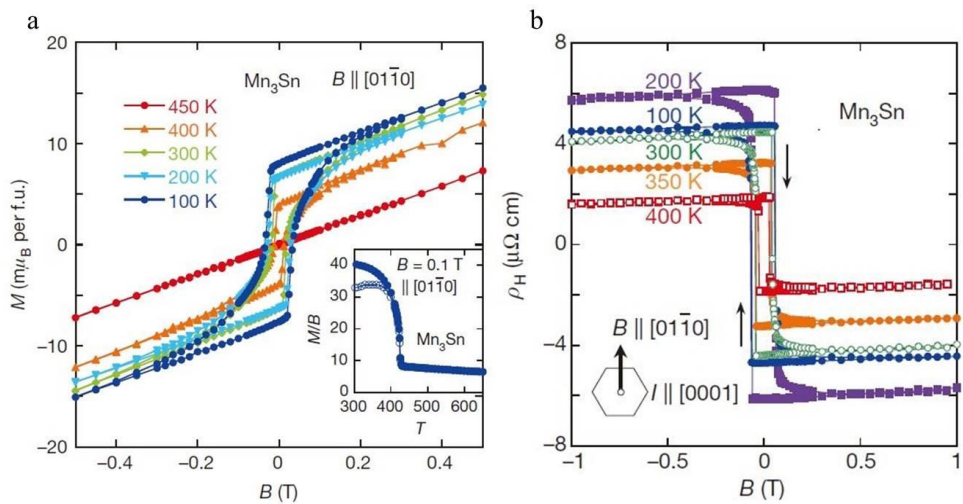

**Fig.8 a** Field-dependent magnetization of $Mn_3Sn$ from 100 to 450 K and temperature-depend magnetization at a magnetic field of 0.1 T (inset). **b** Anomalous Hall resistivity of $Mn_3Sn$ at various temperatures. [37]

On the other hand, Weyl fermions, originating from the linear crossing of non-degenerate bands near the Fermi level in three-dimensional momentum space and usually arising in pairs with adverse chirality due to the structural periodicity, have excited a huge amount of research interest in the condensed matter physics. Negative magnetoresistance could appear when the current flows parallel to the applied magnetic field because of the chiral anomaly which is one of the most important characteristics for Weyl systems. The non-trivial properties of Weyl semimetals quite differ from those in topological insulators in the case of magnetotransport. The Fermi arcs, connecting the two Weyl points, are known as topologically protected zero-energy surface states. The emergence of Weyl fermions requires either time reversal break or inversion symmetry break. The first experimental realization of Weyl fermions was based on TaAs with a non-centrosymmetric structure. However, a time-reversal-break system for observing Weyl fermions had been rare.

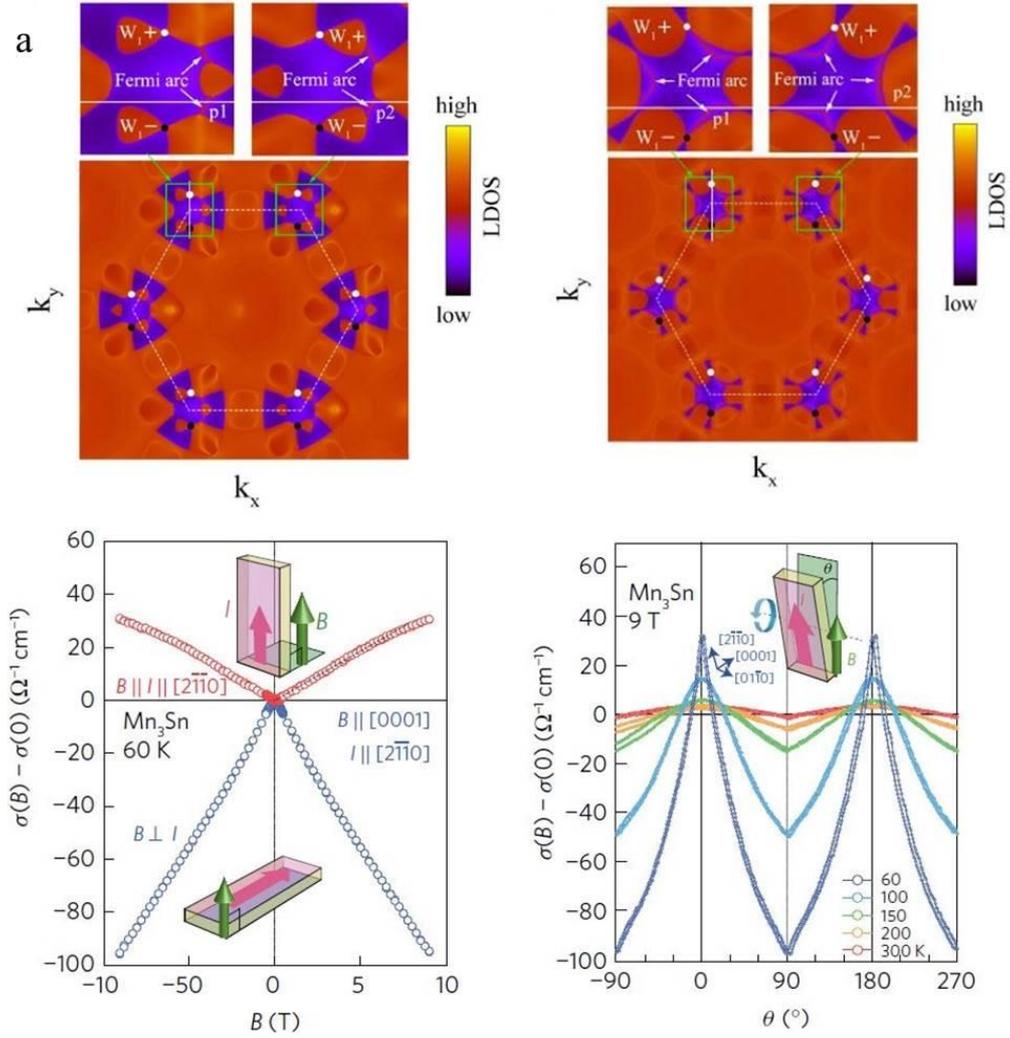

**Fig.9** Fermi surface crossing the $W_1$ Weyl points of **a** $Mn_3Sn$ at $E_F = 86$ meV and **b** $Mn_3Ge$ at $E_F = 55$ meV. [39] **c** Magnetoconductivity measurements as current is applied parallel or perpendicular to the magnetic field. **d** Magnetoconductivity measurements under dependence of angle between the directions of the magnetic field and current. [40]

In 2017, Yang *et al.* [39] discovered the Weyl points in antiferromagnetic $Mn_3Sn$ and $Mn_3Ge$ by *ab initio* band structure calculations, implying the Weyl semimetal state in $Mn_3Sn$ and

Mn$_3$Ge. The nodes of the conduction and valence band crossing near Fermi level could be Weyl points. As seen in Fig. 9a, the Fermi arc connecting the two Weyl points with opposite chirality originates from two neighboring Weyl pairs in Mn$_3$Sn. However, Mn$_3$Ge shows a more complicated Fermi surface and they are divided into three pieces (Fig. 9b). The number of Weyl points is less in Mn$_3$Sn due to the stronger spin-orbit coupling. The Weyl points serving as the source of Berry curvature lead to the anomalous Hall effect in Mn$_3$Sn and Mn$_3$Ge.

Almost at the same time, Kuroda *et al.* [40] provided experimental evidence for the existence of Weyl fermions in Mn$_3$Sn with the assistance of angle-resolved photoemission spectroscopy measurements and density functional theory calculations. Based on the results of magnetotransport measurements, the sign of longitudinal magnetoconductance is positive when the current is applied parallel to the magnetic fields (Fig. 9c). The magnetoconductance varies sharply under the dependence of angle between current and fields (Fig. 9d). The longitudinal magnetoconductance and its anisotropy are consistent with the chiral anomaly which substantiates the exotic Weyl state in the noncollinear antiferromagnetic Mn$_3$Sn.

**2.3 Spin-polarized current and spin Hall effect without spin-orbit coupling**

Spin-polarized current is a powerful tool for realizing various spintronic devices. In the past few years, spin-polarized current has been traditionally achieved in ferromagnetic and ferrimagnetic materials. Nevertheless, antiferromagnetic materials with zero net moment are generally thought to generate zero spin polarization. However, in 2017, Železný *et al.* [41] theoretically discovered the presence of spin-polarized current in noncollinear antiferromagnetic Mn$_3$Sn and Mn$_3$Ir. Spins with a good quantum number in ferromagnets can flow up and down even in the absence of spin-orbit coupling and consequently the current becomes spin polarized. As Fig.10a depicts, in the noncollinear antiferromagnetic materials, the spins of electrons at the Fermi level arrange in distinct directions and the integral spin moment is zero as a consequence of full compensation. However, when an electric field is applied, the electrons redistribute and thus a net current accompanying with a net spin current appears (Fig.10b). In addition, the spin transfer torque induced by the spin-polarized current is able to switch the resistance state in an antiferromagnetic heterostructure which is similar to a magnetic tunnel junction (Fig 10c and d).

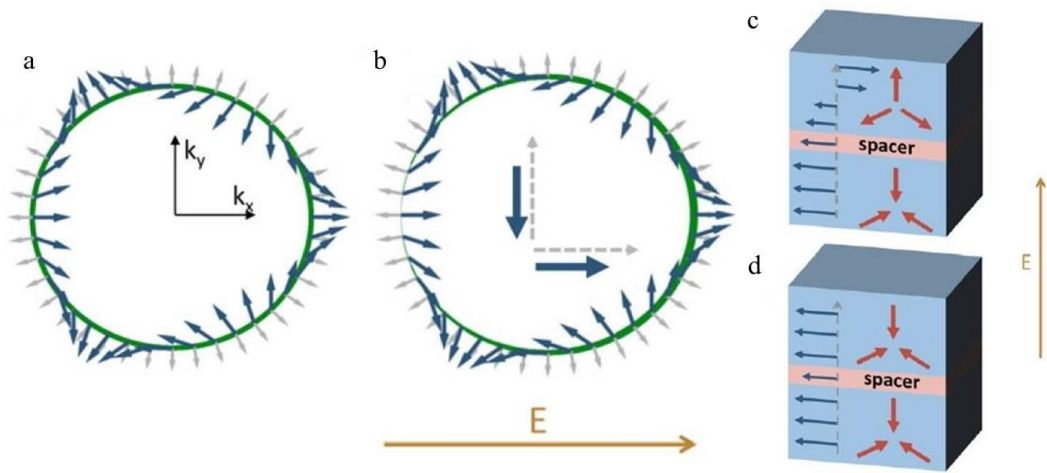

**Fig.10** Illustrations of **a** simplified Fermi level and **b** redistribution of electrons under the applied electric fields at the Fermi level in the noncollinear antiferromagnets. Antiferromagnetic junctions with **c** parallel state and **d** antiparallel state. [41]

When a longitudinal electric current is applied, the electrons with different spins migrate oppositely in the transverse direction and thus the spins accumulate at the edge. Consequently, a spin current is generated and this phenomenon in analogy to the anomalous Hall effect is known as the spin Hall effect. The spin current can be exerted to manipulate magnetic moments, and induce the motion of domain walls. The spin Hall effect is believed to originate from spin-orbit coupling for a long time. In addition, its inverse effect called the inverse spin Hall effect has been extensively utilized to detect the spin current.

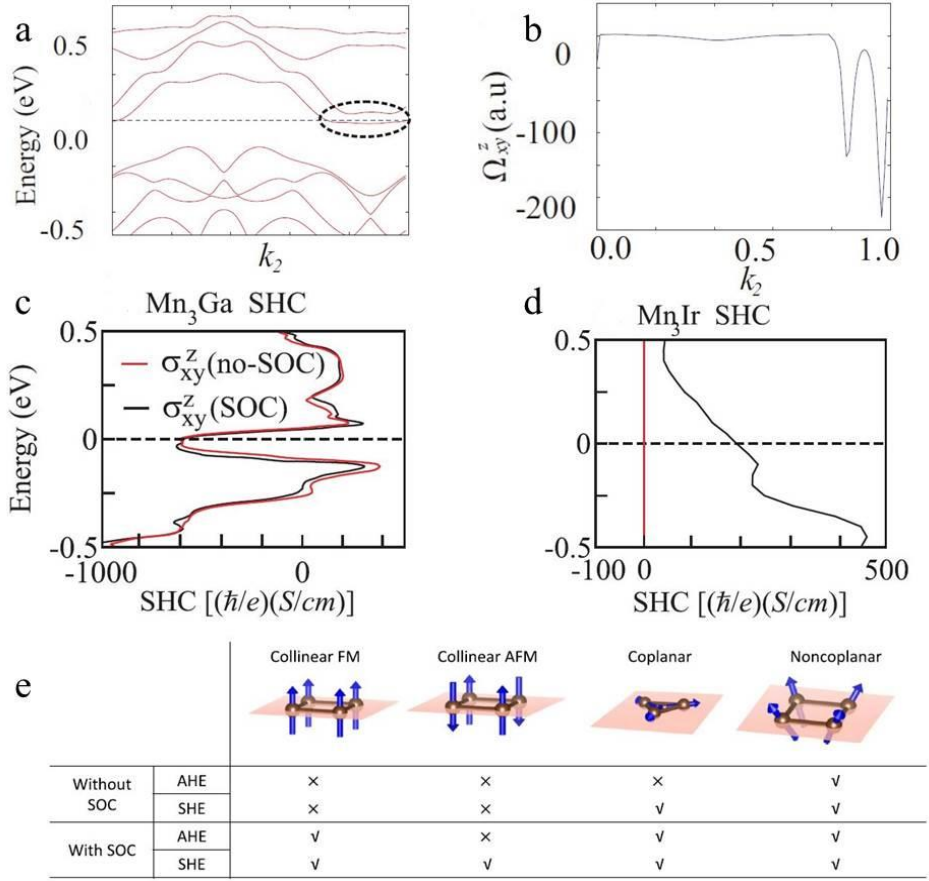

**Fig.11 a** Band structures of $Mn_3Sn$ along $k_2$. **b** Berry curvature evolutions along $k_2$ for $Mn_3Sn$. [42] Energy dependence of spin Hall effect tensor element with spin-orbit coupling and without spin-orbit coupling in **c** $Mn_3Ga$ and **d** $Mn_3Ir$. **e** Summarization of essential symmetry requirements for the spin Hall effect. [43]

In 2017, Zhang *et al.* [42] theoretically found the spin Hall effect in chiral antiferromagnetic compounds $Mn_3X$ ($X$ = Ge, Sn, Ga, Ir, Rh, and Pt). In this work, the Berry curvature reaches maxima at positions where tiny band gaps near the Fermi level occur in the band structures (Fig 11a and b). And this may be the reason for the intrinsic spin Hall effect. Soon afterwards, they revealed that the spin Hall effect in $Mn_3Ge$, $Mn_3Ga$ and $Mn_3Sn$ can exist even in the absence of spin-orbit coupling [43]. As illustrated in Fig. 11c, the effect of spin-orbit coupling on band structure is so weak that the spin Hall effect barely changes. However, it is clear in figure 11d the spin Hall effect vanishes in the Mn$_3$Ir without spin-orbit coupling as a result of the high symmetry with cubic structure. The authors explained that the leading factor to the spin Hall effect is the noncollinear magnetic order. Additionally, spin-orbit coupling can behave in a similar fashion to break the symmetry of spin rotation. More importantly, the basic, essential symmetry requirements for the spin Hall effect were summarized by the authors as shown in Fig. 11e. This work suggests that the spin Hall effect does not necessarily need heavy elements.

## 3. Electric-field control of noncollinear spin structure

In 2016, Nakatani *et al.* [44] switched the creation and annihilation of a stable skyrmion by electric field pulses in a nanometer scale disk with the help of micromagnetic simulation. In this work, the effect is considered to stem from the variation of the magnetic anisotropy induced by the electric field with opposite polarity in the nanodisk. As evident from Fig. 12a, when a negative field is applied, a skyrmion forms instead of the initial single domain. When the electric field is positive, the skyrmion vanishes and degrades into a different single domain. Interestingly, the core direction of the skyrmion also reverses during this process and it will recover after another bipolar pulse. The pulse width is extremely small, ~10 ns, which does not require exact control and an extra magnetic field. The novel and reversible method of modulating the skyrmions can provide potentials in electric-field-control nanoscale devices. Moreover, Ma *et al.* [45] reported that the electric field could tune the creation and motion direction of skyrmion bubbles at room temperature (Fig. 12b), which is of great significance to the ultralow-energy-consumption devices based on skyrmions.

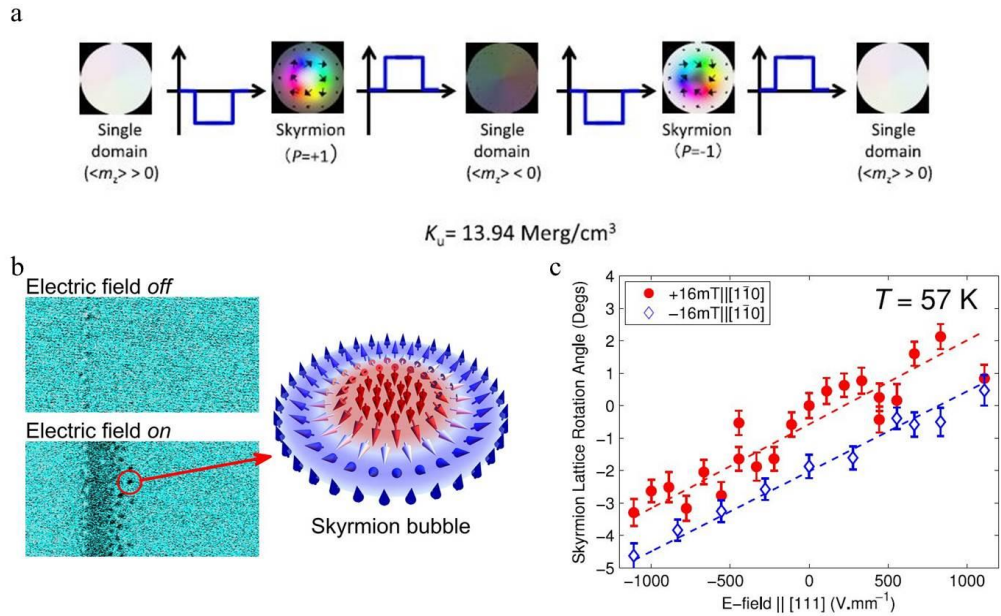

**Fig.12 a** Sketch of evolution of the magnetic states when four electric field pulses are applied successively. [44] **b** Creation and directional motion of skyrmion bubbles under the electric field. [45] **c** The rotation angle of skyrmion lattice under dependence of electric field at 57 K. [46]

White *et al.* [46] demonstrated that the rotation of skyrmions could be manipulated by electric fields in magnetoelectric insulator $Cu_2OSeO_3$ with chiral lattice. The skyrmions rotate around the magnetic field axis upon the application of an electric field and the angle of rotation is linear with the applied electric field (Fig. 12c). Furthermore, the sign of angle is related to the polarity of the electric field instead of the magnetic field. The mechanism of the rotation of skyrmions may be the coupling between the electric field and the local electric dipole at the center of skyrmions lattice. After that, researchers from the same group reported that the electric fields can obviously regulate the expansion and shrinkage of the skyrmion phase stability in $Cu_2OSeO_3$ [47]. Through the small-angle systematic analysis of neutron scattering, Fig. 13a shows that a positive electric field of +5.0 V/μm can expand the skyrmion phase to about twice larger while a negative electric field of −2.5 V/μm shrinks the area of stable phase. The exotic phenomenon in this work is thought to come from the free energy of the skyrmion phase induced by electric fields and magnetoelectric coupling in $Cu_2OSeO_3$ (Fig. 13b). The experimental results are in good agreement with calculations and the optimum parameters provided in this work will assist to enhance the stability of skyrmions.

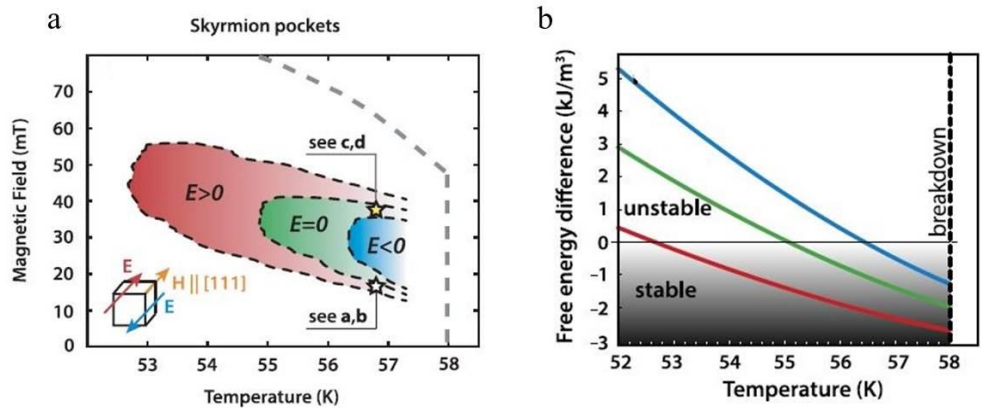

Fig.**13 a** Skymion pockets with different electric fields in phase diagram. **b** Temperature-dependent free energy difference for various electric-field state. [47]

In 2017, Ohuchi *et al.* [48] indicated that the topological Hall effect can be tailored by electric fields in oxide heterostructures constitutive of ferromagnetic $SrRuO_3$ and nonmagnetic $SrIrO_3$. The topological Hall effect is a combination of spin-orbit coupling and Dzyaloshinskii–Moriya interaction caused by symmetry breaking. As seen in Fig. 14a, the topological Hall effect can be remarkably controlled by an electric field at 30 K. The topological Hall resistivity increases to ~60 μΩ·cm under -180 V and decreases to ~20 μΩ·cm when an electric field of +200 V was applied. The peak position at ~0.8 T in the Hall curves is negligibly changed. The phase diagrams (Fig. 14b-d) which depends on temperature and magnetic field present the expansion of the topological Hall region upon the application of negative electric fields. The sign of topological Hall effect stays the same and is independent with the electric-field control. The authors attributed the modulation to the variation of the spin-orbit coupling at the interface of SrRuO$_3$ and SrIrO$_3$ and emphasized the importance of nonmagnetic interlayer SrIrO$_3$ that has large spin-orbit coupling in this work.

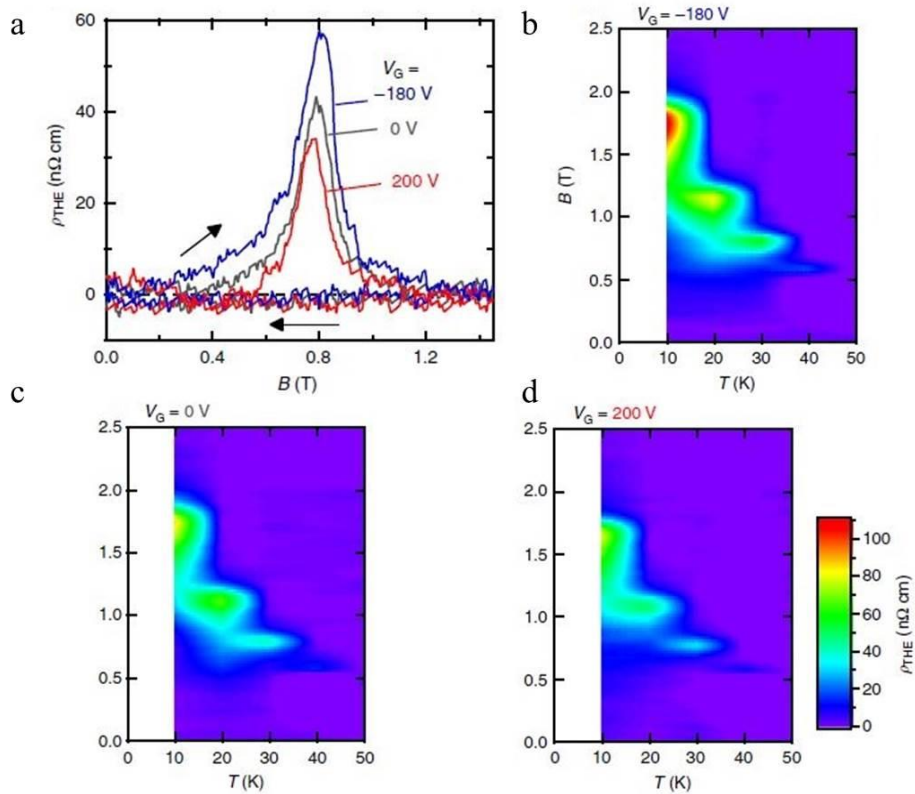

**Fig.14 a** Topological Hall resistivity when gate voltage is -180, 0 and +200 V. Color map of topological Hall resistivity with temperature and magnetic field when the gate voltage is **a** -180, **b** 0 and **c** +200 V. [48]

Most recently, Qin *et al.* [49] realized the electric-field control of topological Hall effect in a single SrRuO$_3$ layer and they applied electric fields by ionic liquid (Fig. 15a). Figure 15b depicts the anomalous Hall effect combined with the topological Hall effect. It can be clearly viewed that the shape of bumps and dips varies via the application of electric fields. In addition, the positive fields enhance the topological Hall effect whereas the negative fields restrains the effect (Fig. 15c), which is in contrast to the work from Ohuchi. The huge disparity in topological Hall resistivity between the two studies suggest that the nonmagnetic SrIrO$_3$ could be crucial to determine the magnitude of Hall resistivity. In this work, the electric potential gradient in the surface of SrRuO$_3$ tunes the spin-orbit coupling thus arouse the change of Dzyaloshinskii–Moriya interaction rather than magnetic exchange energy or the anisotropy energy of the single SrRuO$_3$ layer.

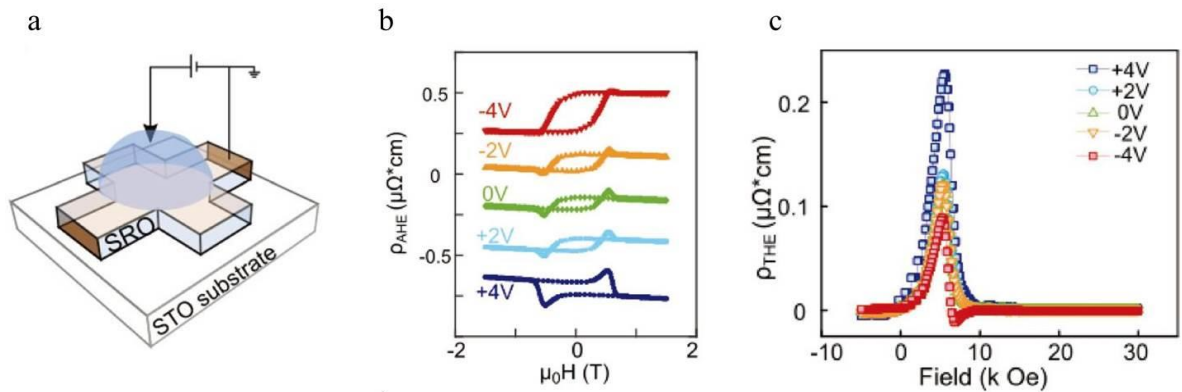

**Fig.15 a** Sketch of Hall bar based on SrRuO$_3$ induced by ionic liquid gating. **b** Hall resistivity at various electric field. **c** Topological Hall resistivity at various electric field of +4, +2, 0, -2 and -4 V. [49]

On the other side, in multiferroic MnWO$_4$, Finger *et al.* [50] demonstrated that the switch of the chiral component by applied electric fields. As given in Fig.16a-d, after cycling the voltage, the chiral radio can always return to the initial value. The width of the hysteresis curves becomes smaller via increasing temperature and the chiral arrangement is completely inversed near the transition temperatures, implying the inversion of the chiral domains. Moreover, the hysteresis curves get wider when the electric voltage reaches -3500 V (Fig. 16e-l) and can be fully reversed by the electric fields in the same way. This work makes the electric-field-control of multiferroic materials as a possible route to information storage device applications.

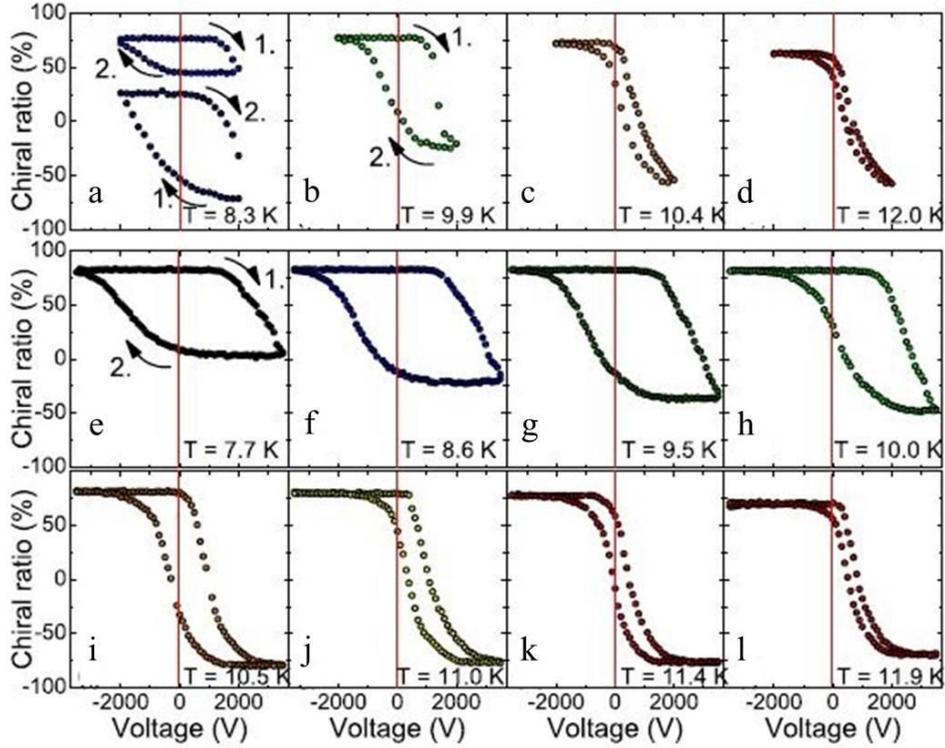

**Fig.16** Hysteresis curves of the chiral ratio under dependence with external electric field at constant temperature after electric-field cooling in **a-d** -2000 and **e-l** -3500 V.

The noncollinear antiferromagnetic $Mn_3Pt$ [51] with a similar triangular spin structure to $Mn_3Ir$ transits to the collinear antiferromagnetic phase at ~360 K (Fig. 17a). At its low-temperature noncollinear phase, it displays a large anomalous Hall resistivity of ~5 $\mu\Omega\cdot cm$ at 10 K and ~2 $\mu\Omega\cdot cm$ at 345 K (Fig. 17b). However, the anomalous Hall effect vanishes at 365 K, consistent with the results of the phase transition. Then the high-quality $Mn_3Pt$ is epitaxially deposited onto ferroelectric $BaTiO_3$ single-crystal substrates. After applying an electric field of 4 $kV\cdot cm^{-1}$, the transition temperature is increased by ~28.5 K as a consequence of the compressive piezoelectric strain of ~0.35 % from $BaTiO_3$ (Fig. 17c). Furthermore, the anomalous Hall effect appears at 360 K under the electric field, suggesting the emergence of the noncollinear spin structure (Fig. 17d). This work proves that the anomalous Hall effect of noncollinear antiferromagnetic materials can be manipulated by electric fields and excited further research in the region.

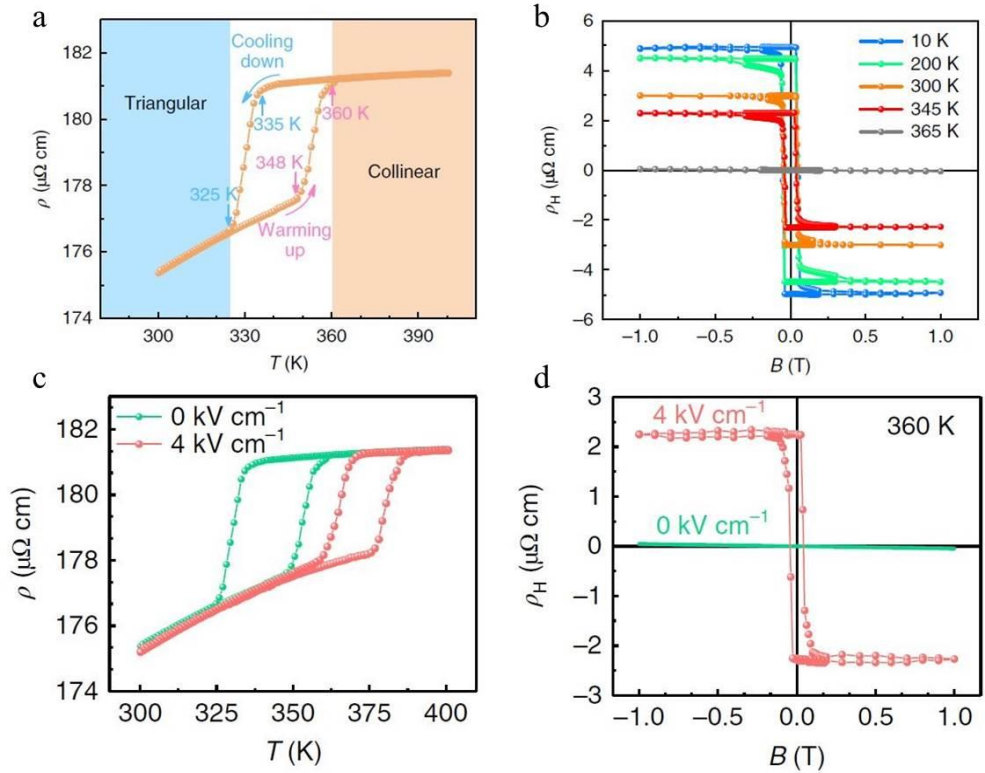

**Fig.17 a** Temperature-dependent resistivity between 300 and 400 K. **b** Hall resistivity at various temperature. **c** $\rho$-T curves under an applied electric field of 0 and 4 kV·cm$^{-1}$ **d** Anomalous Hall effect under an electric field of 0 and 4 kV·cm$^{-1}$ at 360 K. [51]

In our recent work [52], we found the evidence for the presence of the Weyl state in antiferromagnetic Mn$_3$Sn thin films through the parallel negative magnetoresistance (Fig 18a) and anisotropic magnetoresistance (Fig 18b). Then we integrated the Mn$_3$Sn thin films onto piezoelectric 0.72PbMg$_{1/3}$Nb$_{2/3}$O$_3$–0.28PbTiO$_3$ (PMN-PT) single-crystal substrates. Nevertheless, the Mn$_3$Sn films are not stable upon the applied electric fields as the obvious cracks can be generated on the surfaces due to the cracking of ferroelectrics. Note that when we inserted a 100-nm-thick LaAlO$_3$ interlayer between the film and substrate (Fig 18c), the cracks can be restrained even though the electric fields are cycled for many times. The noncollinear spin structure of Mn$_3$Sn is so sensitive to the piezoelectric strain that an electric field of 3.6 kV·cm$^{-1}$ enhances the anomalous Hall resistance from 4.1 mΩ to 11.9 mΩ at zero magnetic field at 150 K (Fig 18d). Moreover, the modulation of the Weyl state could be realized by electric fields as it is closely related to the noncollinear spin structure that breaks the hexagonal symmetry of the Mn$_3$Sn crystal.

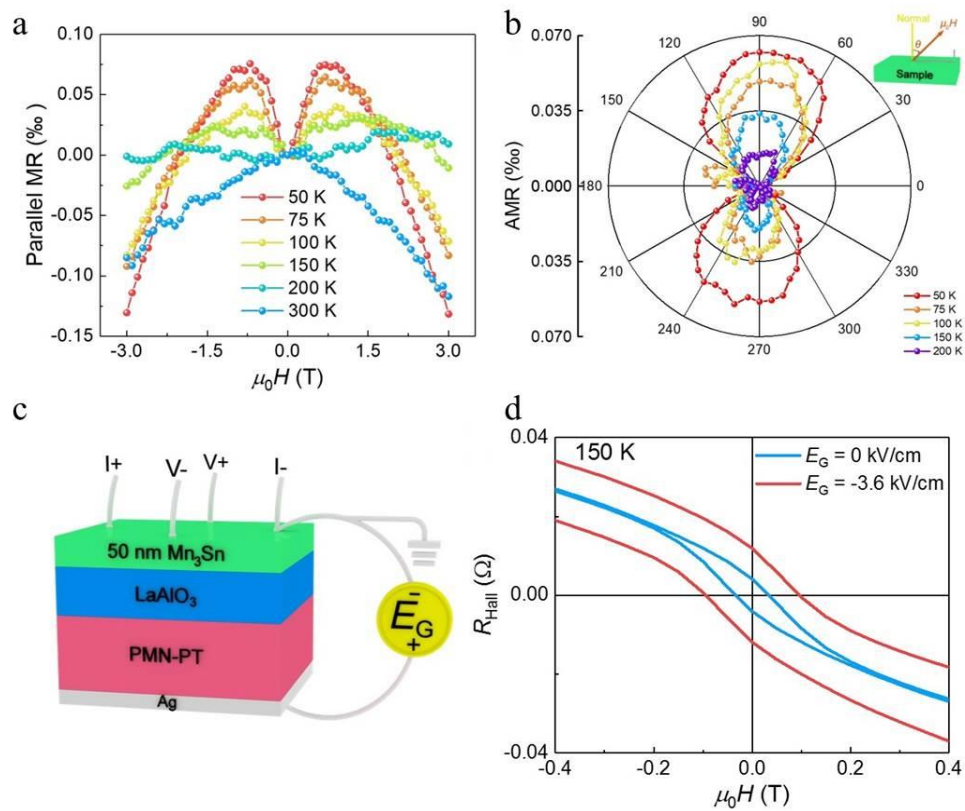

**Fig.18 a** Magnetoresistance measurements at different temperature ranging from 50 to 300 K with the magnetic field applied parallel to the current. **b** Magnetoresistance measurements as a function of angle between the direction of magnetic field and sample. **c** Schematic of a Mn$_3$Sn (50 nm)/LaAlO$_3$ (100 nm)/PMN-PT multiferroic heterostructure and the field-effect measurement geometry. **d** Anomalous Hall effect under an applied field of 0 and -3.6 kV·cm$^{-1}$. [52]

Very interestingly, Lukashev *et al.* [53] demonstrated the theory of piezomagnetic effect in Mn-based antiperovskites with a triangular spin structure akin to Mn$_3$Pt and Mn$_3$Sn. As listed in Fig. 19a, the local magnetic moment in noncollinear Mn$_3$GaN is closely related to the strain.

When the tensile strain is applied, the local magnetic moment rotates in the opposite direction relative to the compressive strain. In addition, the local magnetic moment decreases with compressive strain while it increases with tensile strain. The relationship between net magnetic moment, rotation angle and biaxial strain is summarized in Fig. 19b. This work reveals the variation of spin structure upon the application of piezoelectric strain and can largely assist the research for piezospintronics in noncollinear antiferromagnets.

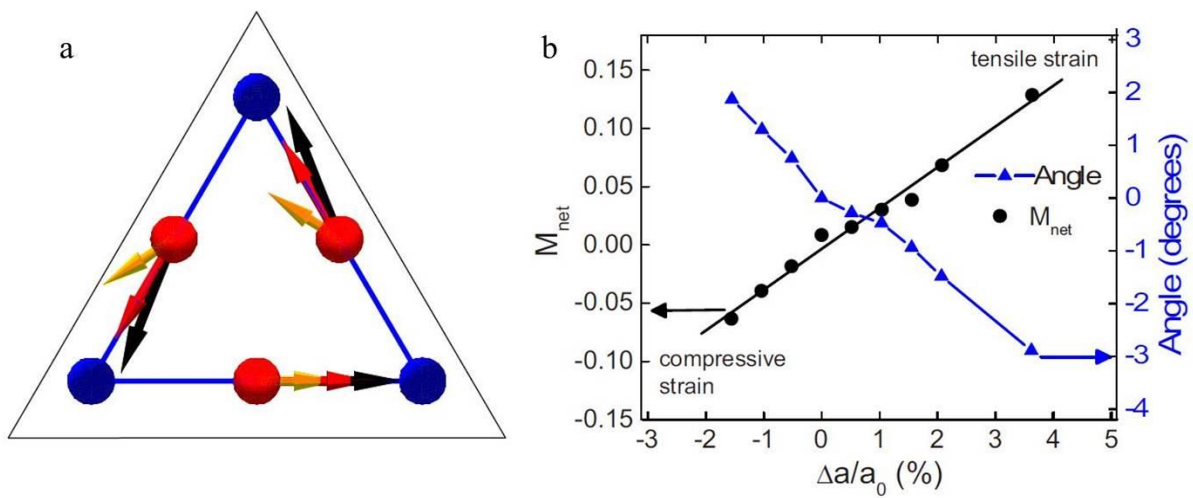

**Fig.19 a** Variation in local magnetic moment of Mn3GaN under the compressive strain (yellow arrow), zero strain (red arrow) and tensile strain (black arrow). **b** Summarization of net magnetization and the rotation angle under dependence with biaxial strain. [53]

## 3. Perspectives

Although the spin-polarized current is predicted to exist in noncollinear antiferromagnetic materials, it has not been experimentally confirmed yet. Based on our recent work on $Mn_3Sn$, the possible Weyl state and the electric-field-control anomalous Hall effect imply that the noncollinear $Mn_3Sn$ may own high spin polarization and become the arena for the study of spin-polarized current. In addition, in the surface/interface system, the Rashba spin-orbit coupling originating from structural inversion asymmetry can be controlled by the gate voltage. In this case, we can apply a proper electric field to offset the Rashba effect and thus build a circumstance in the absence of spin-orbit coupling (Fig. 20). As a result, the spin Hall effect without spin-orbit coupling can be further explored with this approach. Additionally, the experimental detection and intrinsic mechanism of anisotropy magnetic resistance in noncollinear antiferromagnetic materials has been fully elusive and remains an open question. Thus, it will be quite interesting if the anisotropic magnetoresistance effect in noncollinear spintronic materials including ferromagnets and antiferromagnetically coupled materials can be confirmed, which is of great interest for developing tunneling junctions based on noncollinear spintronics for high-density information storage in the future.

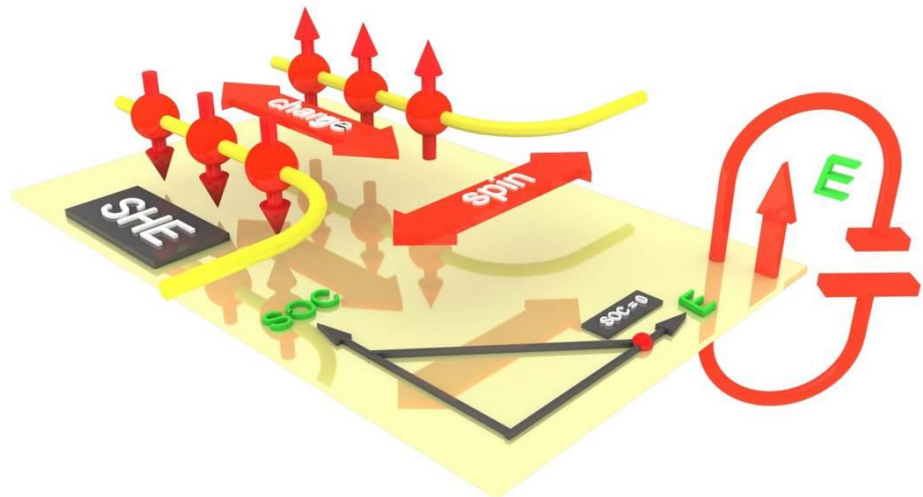

**Fig.20** Sketch of spin Hall effect without spin-orbit coupling under an applied electric field.

In addition, we would like to briefly comment on the magnetic thin film growth on common ferroelectric single-crystal substrates such as PMN-PT and $BaTiO_3$ and relevant electrical measurements, which may be useful for future studies. Based on our research experience, a too high deposition temperature of antiferromagnetic intermetallic thin films in sputter could always lead to ferromagnetic defects such as in MnPt [54] due to the large surface energy difference between intermetallic materials and oxide substrates at high temperatures and the resulting wetting issue [14,55]. Therefore, for fabricating antiferromagnetic thin films, it is better to perform depositions at relatively low temperature, for example, below 450 °C. In addition, ferroelectric substrates are fragile upon cycling periodic electric fields and tend to crack [56]. An effective way to avoiding this issue is to use the unipolar measurement method for ferroelectrics [15,57] or to add a dielectric buffer layer such as $LaAlO_3$ between intermetallic films and ferroelectric substrates [52]. In addition, compared with PMN-PT, $BaTiO_3$ is much more stable at high temperatures similar to $SrTiO_3$ [58-60] and can exhibit large piezoelectric strain but it provides smaller nonvolatile strain [51,61]. Therefore, PMN-PT substrates are better for realizing low-temperature magnetic thin film growth and memory devices while $BaTiO_3$ substrates are more reliable for high-temperature-growth and larger piezoelectric modulation.

It is indeed rather promising to combine noncollinear spintronic materials with an ultrathin ferroelectric oxide thin film to build a complex magnetic/ferroelectric/magnetic heterostructure, which could enable the electric-field control of the tunnel magnetoresistance via either ferroelectric polarization switching [62] or oxygen manipulation [63,64]. Despite of the fact that most of well-studied noncollinear magnetic materials are intermetallic alloys, it is possible to realize the exotic noncollinear spin order in ultrathin magnetic oxide systems such as $LaMnO_3$ [65,66] or $SrRuO_3$ [67] through multiple possible approaches, for example, oxygen octahedral tilting [68,69], direct [70] or long-range [71] exchange coupling, and inversion symmetry break by interfacing a heavy metal [72], which could further become free standing by using a water-dissolvable $Sr_3Al_2O_6$ buffer layer [73] and thus facilitate the transfer of noncollinear oxide based spintronic devices onto silicon.

Lastly, it is interesting to notice that recent studies on antiferromagnetic materials have demonstrated that domain walls can significantly scatter electrons and thus increase resistance by several or even ten percent, which is one or two orders of magnitude larger than the anisotropic magnetoresistance (~0.1%) of antiferromagnetic materials. For example, the work by Kaspar *et al*. [74] reported that the Joule heating induced by a large current pulse can shatter large antiferromagnetic domains into nano-texture domain structures where the domain wall size could be comparable to the domain size. As a result, the largely increased domain wall density leads to an over 10% resistance enhancement in collinear antiferromagnet CuMnAs. In addition, Yu *et al.* [75] also found that the decrease of the domain wall density in collinear antiferromagnet Mn$_2$Au triggered by the spin-flop transition can result in a 1% resistance drop. The role of domain walls in determining the electrical transport properties of noncollinear spin systems could be similar or even greater such as in skyrmion systems.

**Acknowledgements**

Zhiqi Liu acknowledges financial support from the National Natural Science Foundation of China (NSFC; grant numbers 51822101, 51861135104, 51771009 & 11704018).